\documentclass[a4paper,pre,reqno,superscriptaddress, twocolumn, floatfix]{revtex4}

\usepackage{preamble}
\def\numdatasets{24 }
\def\ho{higher-order }

\begin{document}

\author{Bern\'e L. Nortier }
    \email{bln1@st-andrews.ac.uk}
    \affiliation{
    Department of Computer Science, 
    University of St.~Andrews, 
    St.~Andrews KY16, Scotland}
\author{Simon Dobson} 
    \affiliation{
    Department of Computer Science, 
    University of St.~Andrews, 
    St.~Andrews KY16, Scotland}
\author{Federico Battiston}
    \email{battistonf@ceu.edu}
    \affiliation{
    Department of Network and Data Science,  
    Central European University Vienna, 
    Vienna 1100, Austria}

\title{Higher-order shortest paths in hypergraphs}

\begin{abstract}
One of the defining features of complex networks is the connectivity properties that we observe emerging from local interactions. 
Recently, hypergraphs have emerged as a versatile tool to model networks with non-dyadic, \ho interactions. Nevertheless, the connectivity properties of real-world hypergraphs remain largely understudied. 
In this work we introduce path size as a measure to characterise \ho connectivity and quantify the relevance of non-dyadic ties for efficient shortest paths in a diverse set of empirical networks with and without temporal information.
By comparing our results with simple randomised null models, our analysis presents a nuanced picture, suggesting that non-dyadic ties are often central and are vital for system connectivity, while dyadic edges remain essential to connect more peripheral nodes, an effect which is particularly pronounced for time-varying systems. 
Our work contributes to a better understanding of the structural organisation of systems with \ho interactions. 
\end{abstract}

\maketitle

\section*{INTRODUCTION}

Networks, collections of nodes and their interactions, are one of the primary tools used to study complex systems, allowing us to describe collective, emergent system properties which can not be captured by looking at the individual units in isolation~\cite{newman2018networks}.
A typical case is the emergence of global connectivity from local interactions, with many real-world networks characterised by the emergence of large connected components~\cite{molloy1995critical}. 
 {Many such systems display low diameter and are hence dubbed `small-worlds'~\cite{wattsstrogatz1998smallworld}, displaying a relational structure able to support both efficient system-wide communications~\cite{latora2001efficient, estrada2008communicability} and making them easy to navigate~\cite{kleinberg2000navigation}.}
The efficient structural connectivity of these systems has a profound impact on their functionality, \textcolor{black}{for both stochastic and deterministic processes.} The presence of shortcuts is known to influence the speed of contagion and emergence of cooperation~\cite{santos2005epidemic}, \textcolor{black}{impacting also} the ability of a system to synchronise~\cite{barahona2002synchronization}.
For all these reasons, efficiently computing shortest paths in graphs is a problem that has attracted enormous interest in the research community for many decades~\cite{dijkstra2022note}, and it continues to do so.

The analysis of connectivity has also produced relevant insights in the case of temporal networks, where edges are not permanent but created and destroyed over time~\cite{holme2012temporal}. 
Structures such as paths~\cite{wu2014path} and connected components~\cite{nicosia2012components} have been formulated in the setting where network structure evolves over time, as well as numerous measures of centrality, ~\cite{nicosia2013graphmetrics}.
 The addition of a temporal dimension, in particular the presence of non-trivial temporal correlations~\cite{karsai2012universal} and memory~\cite{williams2022shape} among interactions, has stimulated important research on increasingly complex temporal network models, able to reproduce empirical patterns~\cite{perra2012activity, lacasa2022correlations, williams2022non}.
Moreover, since the first observations about epidemic processes~\cite{karsai2011small, rocha2011simulated}, network temporality was found to have interesting consequences also for the network dynamics, influencing the behaviour of individuals across a range of processes~\cite{scholtes2014causality, starnini2017equivalence}.

Networks have classically modelled interactions through links, describing relations between pairs of entities only, even though in many real-world systems interactions simultaneously involve multiple nodes~\cite{battiston2020honetworks}. 
Recently, a wide variety of structural descriptors have been extended to account for the presence of such \ho interactions, including algorithms for motif discovery~\cite{lotito2022homotifs} and analysis~\cite{mann2021randomgraphs}, community detection~\cite{eriksson2022flow, ruggeri2023community}, centrality measures~\cite{benson2019three}, as well as network filtering~\cite{musciotto2021detecting} and reconstruction~\cite{young2021hypergraph} procedures. Non-dyadic interactions generate new dynamical behaviours and collective phenomena~\cite{battiston2021physics, bick2023higher}, from contagion~\cite{iacopini2019simplicial, dearruda2020social} to synchronisation~\cite{skardal2020higher, millan2020explosive, zhang2023higher} and evolutionary games~\cite{alvarez2021evolutionary, civilini2024explosive}. More recently, temporality has also been considered in \ho networks, from burstiness~\cite{cencetti2021temporal} and temporal-topological correlations~\cite{ceria2023temporal}, to  {structural models with~\cite{gallo2024higher, iacopini2024temporal}  and without~\cite{ petri2018simplicial, di2024percolation} memory}.

Despite these advances, characterising shortest paths and connectivity in systems with \ho interactions remains an open problem. Recently, efforts have been devoted to characterise the concepts of distance~\cite{vasilyeva2023distancesho} and walks~\cite{aksoy2020hypernetwork} in networks with non-dyadic ties, as well as proposing efficient algorithms to extract shortest paths in hypergraphs~\cite{gao2014hoshortestpath},  {and randomise hypergraphs preserving shortest path lengths~\cite{najajima2022randomising}. However, all these are analyses limited to static systems.

In this work, we provide the first systematic investigation of the effect of \ho interactions on system connectivity across a variety of real-world datasets  {of social interactions.}
We quantify the contribution of non-dyadic ties for shortest path length, computing length distribution, average path size and the fraction of purely dyadic segments in each path, and show that all such features are compatible with a simple null-model which preserves the \ho degree distribution. 
Next, we consider time-varying interactions, and extend the concept of \ho path and components to temporal hypergraphs. 
Our analysis of multiple real-world systems reveals the crucial role of \ho interactions to ensure efficient connectivity in temporal \ho networks, characterises differences between topologically shortest and temporally fastest paths, and shows how the observed non-trivial empirical patterns can not be reproduced with simple randomised null-models. Our work provides new insights into the connectivity properties and the structural organisation of real-world hypergraphs, both for static and time-varying systems.

\section*{RESULTS}

\subsection*{Data}
To study shortest paths in temporal hypergraphs, we collected \numdatasets publicly available datasets of real-world temporal systems with \ho interactions.  {Specifically, our study focuses on hypergraphs that describe social interactions} from a broad range of domains such as households and schools, hospitals, workspaces and conferences, as well as political interactions between individuals in Congress and even contact data from baboons. 

\begin{table}[hb!]
    \begin{tabular}{lcccccc}
    \hline
    \textbf{Dataset} & {$V$} & {$E_{HO}$} & {$E_{DY}$} & {$|e_{\max}^{HO}|$} & {$T$} & {$dt$} \\ 
    \hline
    Copenhagen   & 692  & 449413 & 38088 & 21 & 8064  & 300s \\
    Elem1 & 339  & 26543  & 14184 & 8  & 2244  & 20s  \\
    F\&F: 2010-08  & 44   & 436  & 163   & 9  & 66  & 8h  \\
    F\&F: 2010-09  & 58   & 1155   & 349   & 11 & 90  & 8h  \\
    F\&F: 2010-10  & 128  & 4832   & 1293  & 13 & 90  & 8h  \\
    F\&F: 2010-11  & 126  & 6094   & 1504  & 20 & 90  & 8h  \\
    F\&F: 2010-12  & 121  & 4529   & 1091  & 21 & 90  & 8h  \\
    F\&F: 2011-01  & 118  & 4070   & 1056  & 11 & 90  & 8h  \\
    F\&F: 2011-02  & 117  & 6340   & 1268  & 20 & 90  & 8h  \\
    F\&F: 2011-03  & 112  & 6529   & 1330  & 10 & 90  & 8h  \\
    F\&F: 2011-04  & 112  & 6787   & 1320  & 16 & 90  & 8h  \\
    F\&F: 2011-05  & 97   & 1257   & 364   & 8  & 18  & 8h  \\
    HS11  & 126  & 3642   & 151   & 44 & 610   & 20s \\
    HS12  & 180  & 7458   & 378   & 46 & 1321  & 20s \\
    InVS13  & 95   & 10502  & 2342  & 43 & 20129 & 20s  \\
    InVS15  & 219  & 35764  & 7846  & 71 & 21536 & 20s \\
    Kenyan  & 75   & 972  & 272   & 11 & 42  & 1hr   \\
    LH10  & 76   & 1854   & 1093  & 5  & 7639  & 20s  \\
    LyonSchool   & 242  & 12704  & 7748  & 5  & 3100  & 20s  \\
    Malawi  & 86   & 14051  & 194   & 24 & 1127  & 20s \\
    Mid1  & 591  & 76062  & 49400 & 10 & 2505  & 20s  \\
    SFHH  & 403  & 10541  & 8268  & 9  & 3509  & 1hr   \\
    Thiers13   & 327  & 7818   & 5498  & 5  & 7375  & 1hr   \\
    Congress Bills & 1718 & 82873  & 13845 & 25 & 5305  & 1 months   \\
    \hline
    \end{tabular}
    \caption{ \justifying 
    Summary statistics of real-world temporal hypergraphs. {$V$} indicates the number of nodes, {$E_{HO}$} denotes the number of \ho interactions and {$E_{DY}$} the number of purely dyadic interactions in the static system, and {$|e_{HO}^{max}|$} indicates the maximum size of a \ho interaction. {$T$} measures the number of total timestamps and {$dt$} the interval between successive timestamps in the temporal hypergraphs.}
    \label{tbl:summstats}
\end{table}

We provide relevant summary statistics for the static versions of each dataset in Table~\ref{tbl:summstats}, alongside the amount and resolution of successive timestamps for the temporal networks. A fuller description of the datasets is provided in Methods.

\subsection*{Static hypergraphs}\label{sec:static}

In simple graphs, relational information is described by nodes and edges, which encode pairwise interactions between pairs of nodes only. 
 {A \textit{path} between 2 nodes $i$ and $j$ is an ordered, non-repeating sequence of pairwise edges} 
Similarly, a hypergraph is a collection of nodes and hyperedges which encode interactions among an arbitrary number of nodes.  {Hyperedges are called \textit{pure pairwise} or \textit{dyadic} when they correspond to the classic notion of an edge by connecting only 2 nodes and are distinguished from \textit{truly \ho} interactions where 3 or more nodes participate in an interaction.} The \textit{size} of a hyperedge is defined as the number of nodes contained therein.
 {A \textit{hyperpath} or a \textit{\ho path} between 2 nodes in a hypergraph is defined as an ordered, non-repeating sequence of hyperedges, where each subsequent step occurs between nodes within the same hyperedge.}

 {The \textit{length} of a \ho path between any two nodes $i$ and $j$, $\ell_{ij}$, is defined as the number of edges traversed.}
If no such path exists between 2 nodes, we set $\ell_{ij}=\infty$. Two nodes are \textit{connected} if a path of finite length exists between them. 
While a pair of nodes can be connected by multiple distinct paths, the \textit{shortest path} between any two nodes in a static network is defined as a path of minimum length, which might not necessarily be unique.  {Technical details regarding the exact computation of such hyperpaths are provided in the relevant section of Methods.}
 
\begin{figure}[hb]
    \begin{subfigure}{\linewidth}
        \includegraphics[width=0.65\linewidth]{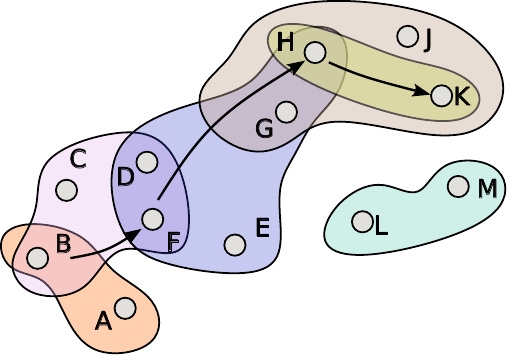}
    \end{subfigure}%
    \caption{\justifying 
     {A shortest hyperpath} between nodes B and K has length $\ell=3$ and average size $\langle{s}\rangle=(4+5+2)/3$.
    }
    \label{fig:1}
\end{figure}

\begin{figure*}[ht]
    \begin{subfigure}{\textwidth}
        \includegraphics[width=\linewidth]{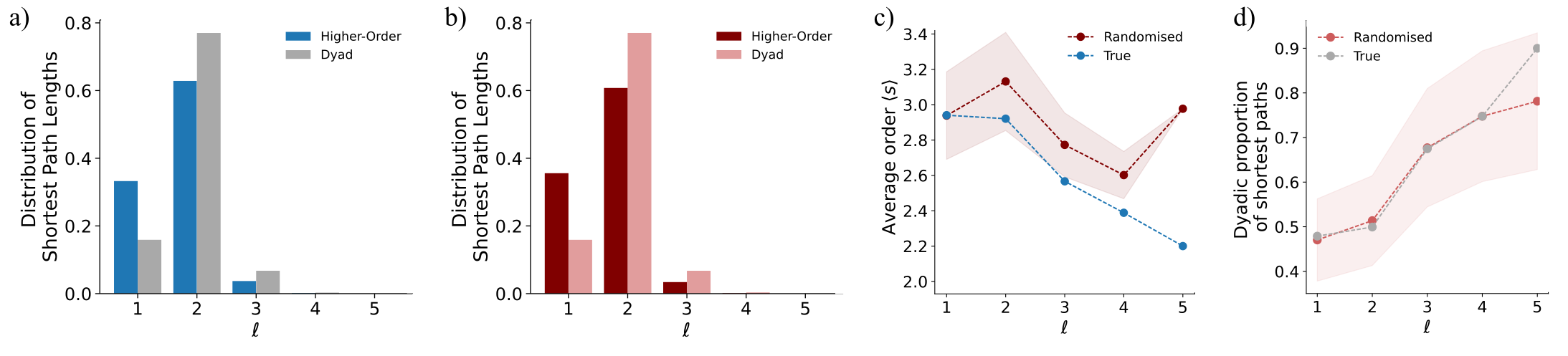}
    \end{subfigure}%
    \caption{ \justifying 
    Shortest paths in the static hypergraph of the Copenhagen study.
    Path-length distributions of \ho and purely dyadic paths for hypergraphs of the Copenhagen dataset for (a) the empirical data and (b) a randomised null-model. 
    (c) Average interaction size $\langle{s}\rangle$ of the \ho path as a function of the path length for empirical (blue) and randomised (dark red) data.
    (d) Average fraction of each path which is purely dyadic as a function of the path length for empirical (grey) and randomised (light red) data.  
    Shaded areas represent a 97.5\% confidence interval obtained over 100 randomisations.}
    \label{fig:2}
\end{figure*}

Clearly, hyperpaths in \ho networks will traverse hyperedges of varying sizes, where each segment of a hyperpath traverses a hyperedge of a potentially different size. As such, one may attach to each individual segment of a hyperpath its associated \textit{size} $s$, defined as the size of the hyperedge that is traversed during that particular step. In some cases, such a step may be contained in multiple hyperedges, in which case we choose the hyperedge of smallest size to define its size.  {Results concerning alternatives, such as selecting the mean or maximum in the case multiple hyperedges, and the frequency with which they occur are reported in Section 3 of the Supplementary Materials. Taken together, our analyses show that different strategies with regard to the hyperedge to traverse in case of redundancies do not induce a qualitative difference in our findings.}
Paths in \ho networks contain rich information and one may calculate the average size $\langle{s}\rangle$ of a path as the average size across traversed segments. 
 {For example, the average size of a path allows one to determine how much of a path makes use of dyads as opposed to genuinely \ho ($s\geq3$) interactions.} 
The minimum size of a path is $\langle{s}\rangle=2$, which occurs when the hyperpath consists of purely pairwise interactions.

To illustrate these concepts, we consider the hypergraph in Figure~\ref{fig:1}. 
 {A shortest hyperpath} between nodes B and K has length $\ell=3$. One of these is the hyperpath (B, F, H, K), where the segments (B, F), (F, H), and (H, K) have sizes 4, 5 and 2.
The average size of such a hyperpath is then $\langle{s}\rangle=11/3$. We note that the segment (H$\to$K) is also contained in a hyperedge of size four but that we select the hyperedge of minimum size to assign a size. Importantly, other shortest hyperpaths of the same length can exist. For instance, here the hyperpath (B, D, H, K) also has the same $\langle{s}\rangle=11/3$. Another of such shortest hyperpaths is (B, D, G, K) which has a higher average size $\langle{s}\rangle=13/3$.

We are interested in studying the \ho organisation of shortest paths and the extent to which group interactions contribute to the forming of these paths in real-world systems.

We begin by considering static hypergraphs where we neglect information about the temporal nature of the interactions (see Methods for details). As an illustrative example, in Figure~\ref{fig:2}, we show results for the social hypergraph from the well-known Copenhagen network study, which describes social interactions over 4 weeks between 576 university students on a university campus.  {Results of analyses on all other datasets are provided in the Supplementary Materials Section 2 and 3.}

In Figure~\ref{fig:2}a, we plot the distribution of shortest path lengths for \ho paths in blue. We compare it with the distribution of path lengths of purely dyadic paths, obtained when taking considering only dyadic interactions, shown in grey.  
When \ho interactions are not considered,
we observe a distribution shift to the right, as such paths take longer to reach the same target as their \ho counterparts.
Interestingly, the same distributions of \ho and dyadic path lengths are accurately reproduced by randomising the data with a \ho configuration model (see Methods for details), as shown in Figure~\ref{fig:2}b.

Next, in Figure~\ref{fig:2}c, we study the average size  $\langle{s}\rangle$ of shortest paths between connected nodes in the \ho network as a function of their length $\ell$. We observe that  {shortest paths} which are longer also tend to be those with a lower average size. 
This suggests that \ho interactions are often central in the system, in agreement with recent work on hypercores~\cite{mancastroppa2024hypercores} while dyadic edges remain crucial to connect more peripheral nodes.  {Due to the small sample (2) of maximum-length paths, the randomisation does not reflect the downward trend at all lengths.} To further corroborate our intuition, in Figure~\ref{fig:2}d we plot the fraction of a shortest path that consists of purely dyadic segments, grouped by the length of the path itself. As expected, the curve is increasing, and such a behaviour is again well reproduced in the randomised system. 

\begin{figure}[ht!]
    \begin{subfigure}{\linewidth}
        \includegraphics[width=\linewidth]{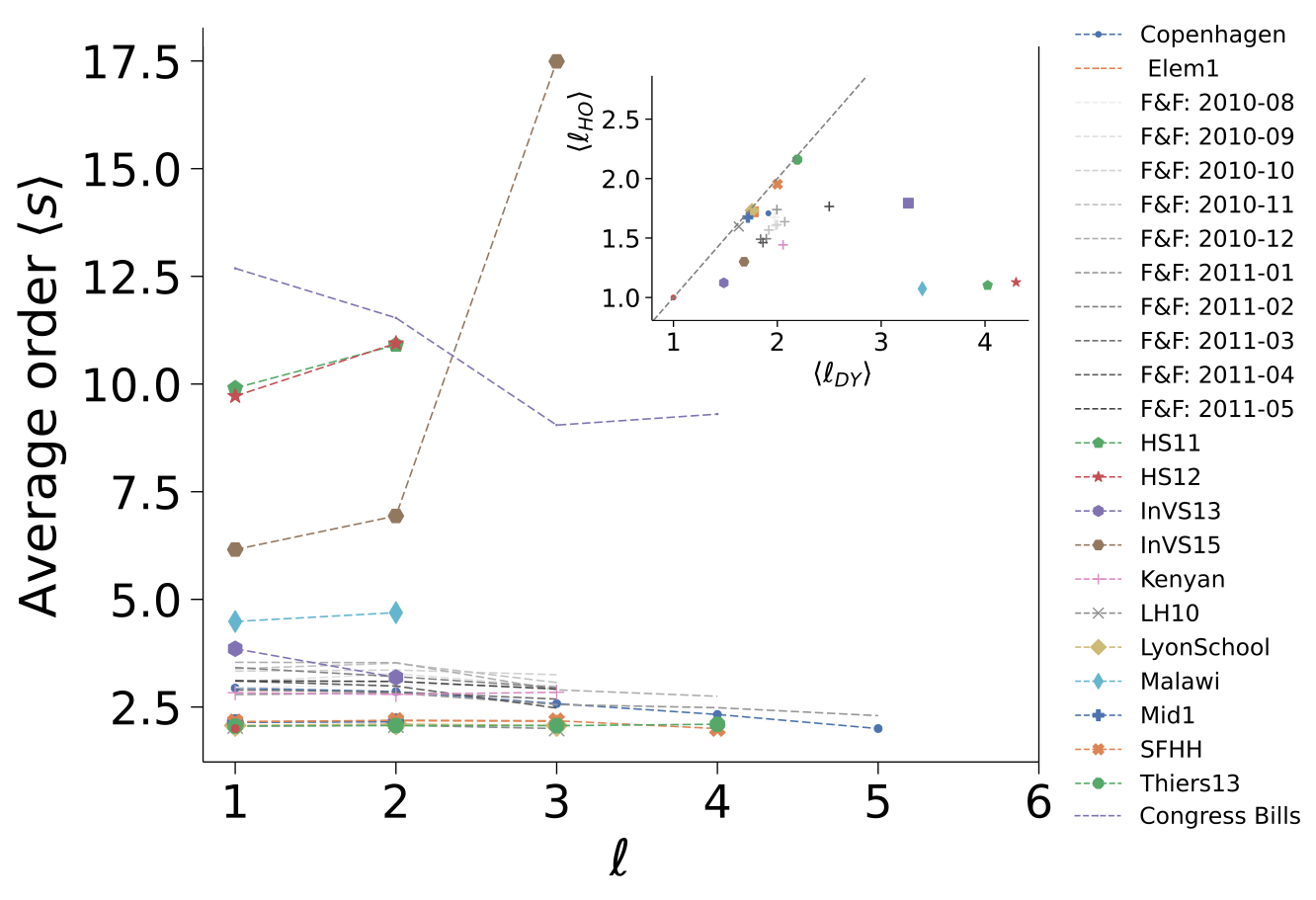}
    \end{subfigure}%
    \caption{ \justifying   
    Summary statistics for shortest paths in static hypergraphs.
    Average size $\langle{s}\rangle$ as a function of the path length across \numdatasets datasets describing social interactions. 
    (Inset) Average shortest \ho path length $\langle \ell_{HO}\rangle$ against the average shortest dyadic path length $\langle \ell_{DY}\rangle$ for all datasets. 
    }
    \label{fig:3}
\end{figure}

\begin{figure*}[ht!]
    \begin{subfigure}{\textwidth}
    \includegraphics[width=0.9\linewidth]{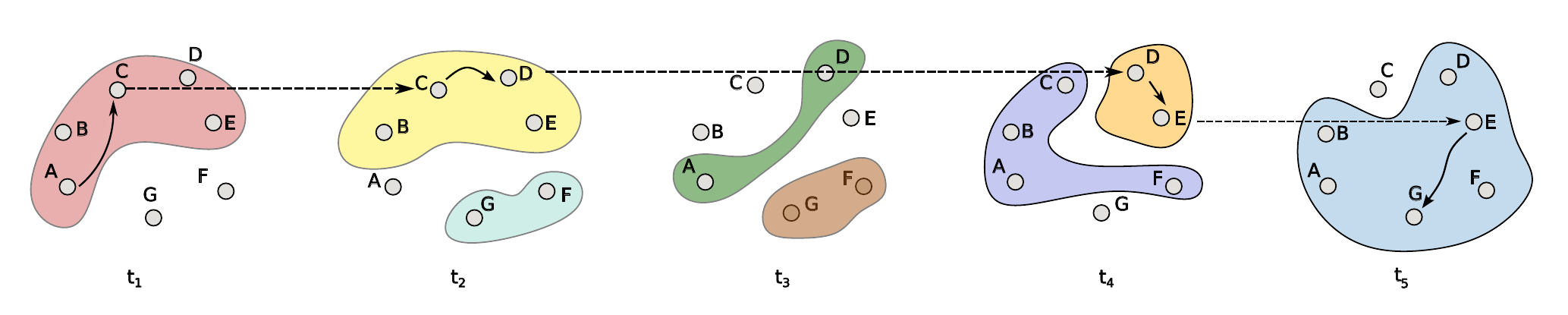}
    \end{subfigure}%
    \caption{ \justifying
        Example of a temporal path travelling from node A at $t_1$ to node G at $t_5$. The path has time duration $d=5$ and path topological length $\ell=4$.
    }
    \label{fig:4}
\end{figure*}

In Figure~\ref{fig:3} we extend the analysis of \ho shortest paths to all our datasets. We observe that the average size $\langle{s}\rangle$ as a function of $\ell$ decreases in all systems, except for a dataset about social interactions in an office.  {We hypothesize that finite-size effects play a role for this workplace data, as the number of paths of length 3 are three orders of magnitude less than the other path length counts.} 

 {The inset of} Figure~\ref{fig:3} shows for all datasets the difference in average path length for \ho and  {purely dyadic} paths.
As expected, those systems which have $\langle{s}\rangle \approx 2$ in the main figure for all lengths are those which fall closest to the line $x=y$ in the inset.
Finally, we observe that the lengths of shortest paths in all datasets are very short, never having a length of more than 5, in agreement with the small-world nature of complex networks~\cite{wattsstrogatz1998smallworld}.

In summary, while \ho interactions are important to ensure connectivity in static hypergraphs, in such a simple scenario most shortest path features can be reproduced by preserving the \ho degree distribution. As we will see in the next section, a much richer picture emerges once we move beyond static systems and expand our analysis to consider the role of time.


\begin{figure*}[ht]
    \centering
    \begin{subfigure}{\linewidth}
        \includegraphics[width=\linewidth]{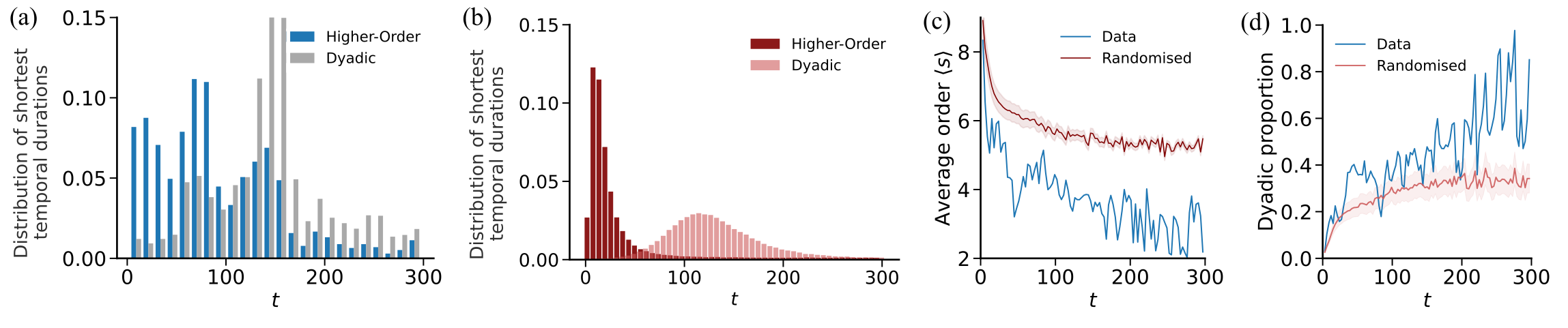}
    \end{subfigure}%
    \caption{ \justifying
        Fastest paths in the temporal hypergraph of the Copenhagen study.
    Path-duration distributions of \ho and purely dyadic paths for hypergraphs of the Copenhagen  dataset for (a) the empirical data and (b) a randomised null-model. 
    (c)  {Average temporal} interaction size $\langle{s}\rangle$ of the \ho path as a function of the path duration for empirical (blue) and randomised (dark red) data.
    (d) Average fraction of each path which is purely dyadic as a function of the path duration for empirical (grey) and randomised (light red) data.  
    Shaded areas represent a 97.5\% confidence interval obtained over 100 randomisations.
    }
    \label{fig:5}
\end{figure*}

\subsection*{Temporal hypergraphs}\label{sec:temp}

We model the temporal evolution of a networked system with a temporal network. A temporal network is a sequence of consecutive time-stamped static graphs, or alternatively a collection of nodes and a collection of timestamped edges. 
 \textcolor{black}{At any time $t$ hyperlinks can be either active or inactive.} 

 {A \textit{temporal path} is an ordered sequence of successive (edge, activation time) pairs, where consecutive pairs' timestamps must be increasing.}  
An important consequence of temporality is that 
paths are not symmetric (even in the case of undirected graphs). In other words, the existence of a path from  $i$ to $j$ does not imply the existence of a path from $j$ to $i$ and in general, even when both exist, they might be different and have different lengths. 
In this work, we assume that a message cannot be propagated further than its direct neighbours within any single snapshot (although the alternative makes sense when timestamps are at longer intervals), 
 and that a node can `retain' a message across multiple snapshots, following the approach of Ref.~\cite{nicosia2013graphmetrics}.

The concept of length in the temporal setting may be defined in 2 complementary ways. 
  The \textit{temporal duration}, denoted $d$, of a temporal path from node $i$ to node $j$ is defined as the time elapsed between the timestamp at which node $i$ is first active and the timestamp at which node $j$ is active. 
  By contrast, the \textit{length} or step-count, denoted $\ell$, of a temporal path from nodes $i$ to node $j$ is defined as the total number of temporal edges traversed.

Shortest temporal paths can therefore be either those that have minimum duration ($d_{\min}$), or minimum length ($\ell_{\min}$). 
 {The \textit{fastest path}~(of shortest duration) between nodes $i$ and $j$ is the temporal path of minimum duration, and is set to infinity if such a path does not exist. The \textit{shortest path} (with the fewest steps) counts the number of steps,} and is hence analogous to the static case.

When a system describes time-varying interactions between groups of nodes, we make use of temporal hypergraphs. A \textit{temporal hypergraph} is a sequence of static hypergraphs, indexed by time. The two notions of temporal length (duration and step count) can be defined as in the pairwise case. 
In Figure~\ref{fig:4}, a simple temporal hypergraph illustrates how such measures can differ. A temporal path  {consisting of timestamped links 
$(A\to{C},t_1), (C\to{D},t_2), (D\to{E},t_4), (E\to{G},t_5)$ has a temporal length of 5 units, but a topological length of 4} links. In our temporal analysis, we mainly focus on fastest paths, although we present some results for both fastest and shortest in Section~4 of the Supplementary Materials.

We investigate temporal connectivity and fastest paths in various real-world systems to discover how much of the system's connectivity is due to  {temporal} \ho interactions. Similarly to the static case, in Figure~\ref{fig:5}, we begin our exploration of the temporal connectivity properties of real-world systems by focusing on the Copenhagen dataset.  {As before, all analysis for the remaining datasets are available in Sections~2 and~3 of the Supplementary Materials, and we mention here only that similar trends are observable across the other datasets.} In Figure~\ref{fig:5}a we plot side-by-side the distribution of durations of fastest paths across all  {temporal} interaction sizes in blue, and for only purely pairwise temporal interactions in grey. 
We note that the duration of such paths can be very long, even well above 100 time units.
When only pairwise interactions are considered, there is a distinct distribution shift to the right of temporal path lengths. Such an effect of \ho interactions is much greater than what we observed for the path length in static hypergraphs, making explicit the very relevant role of non-dyadic connections in determining the connectivity properties of temporal systems.

In Figure~\ref{fig:5}b we plot the same distribution of durations for a randomisation of the data where hyperedges' time-information is shuffled but the underlying static graph is kept constant. 
As shown, such a randomisation fails to reproduce the distribution of durations of the actual system, as washing out temporal correlations among hyperedges makes the typical duration of fastest paths in the null model much shorter and narrower than in the  {real-world} data.

Next, in Figure~\ref{fig:5}c we plot the average size $\langle{s}\rangle$ of fastest paths between connected nodes in the \ho network, grouped by their duration, for both real and randomised data. 
As in the static case, we observe that paths that last longer also tend to have a lower average size in both the true and randomised networks. 
Faster paths traverse more \ho temporal interactions. In contrast, paths with longer duration use more dyadic edges, as illustrated in Figure~\ref{fig:5}d, which shows the average fraction of fastest paths composed of pure dyadic temporal interactions, grouped by path duration. 
While the null model qualitatively reproduces the observed trends, we observe that fastest paths in the randomised data systematically underestimate the usage of dyadic edges in temporal paths. 

In Figure~\ref{fig:6}a we broaden our investigation to consider fastest paths for all datasets in our collection and plot $\langle{s}\rangle$ against $t$, the number of timesteps across all datasets. Again, we observe again the same downward trend of decreasing average size with increasing path duration which is even more pronounced when plotting the topological length of fastest paths, and an increasing dyadic proportion for longer path lengths (last two not shown here). \textcolor{black}{In Figure~\ref{fig:6}b, we calculate the proportion of node-pairs in each dataset that remain connected when dyadic edges are removed, i.e. when they are only allowed to traverse temporal interactions involving groups.} While there is variability across real-world systems, we observe that in many cases a high percentage of paths disappear when only  {pure} dyadic segments are considered, highlighting the importance of \ho interactions for the global architecture of the system. 
For those nodes that remained connected even when only dyadic temporal interactions were allowed, in Figure~\ref{fig:6}c, we visualise how much longer on average a dyadic temporal path will tend to be than its \ho counterpart. We note that in a some cases such differences can be very large, such as for the InVS13 and the Baboons datasets. 

\begin{table}[h!]
\begin{tabular}{l|cc|cc}
\hline
\multirow{2}{*}{\textbf{Datasets}} & \multicolumn{2}{c|}{{$NCC$}} & \multicolumn{2}{c}{{$|LCC|$}} \\
\cline{2-5}
 & Static & Temporal & Static & Temporal \\
\hline
Malawi  & 1 & 30720 & 1.00  & 0.35 \\
LH10  & 1 & 4   & 1.00  & 0.13 \\
F\&F 2010-10 & 1 & 96  & 1.00 & 0.27 \\
F\&F 2010-08 & 2 & 992   & 0.95  & 0.39 \\
F\&F 2010-09 & 1 & 128   & 1.00  & 0.31 \\
Kenyan  & 3 & 692   & 0.41  & 0.17 \\
Thiers13   & 1 & 50  & 1.00 & 0.01  \\
\hline
\end{tabular}
\caption{ \justifying
Connected components in static and temporal hypergraphs. 
The first column presents the number of connected and temporally strongly connected components (NCC) for static and temporal versions of the dataset. The second column gives the relative size of the largest connected and strongly temporally connected components ($|LCC|$) when counting included nodes.
}
\label{tbl:conncomps}
\end{table}

Path temporality also has consequences for the concept of connected components. 
 Following Ref.~\cite{nicosia2012components}, we define two nodes as \textit{strongly temporally connected} if there is a temporal path from $i$ to $j$ and also vice versa. A \textit{temporal strongly connected component} is the set of vertices for which any participating node can both reach and be reached by all other nodes in the set. 
 To quantify the reachability of nodes in the systems, we study the temporal connectivity of the datasets in Table~\ref{tbl:conncomps} by calculating the number of the strongly connected temporal components and the size of the largest. We compare this with the number and largest size of connected components in the static network. The temporal calculation is known to be an NP-complete problem~\cite{nicosia2013graphmetrics} which is only computationally feasible for smaller graphs, so we present here results only for those networks that we were could compute. We observe that the static hypergraph for nearly all datasets has a single giant connected component containing all nodes. In contrast, the temporal systems will be much more fractured, with components that are more numerous and smaller in size.

\begin{figure}[ht]
    \begin{subfigure}{\linewidth}
   \includegraphics[width=\linewidth]{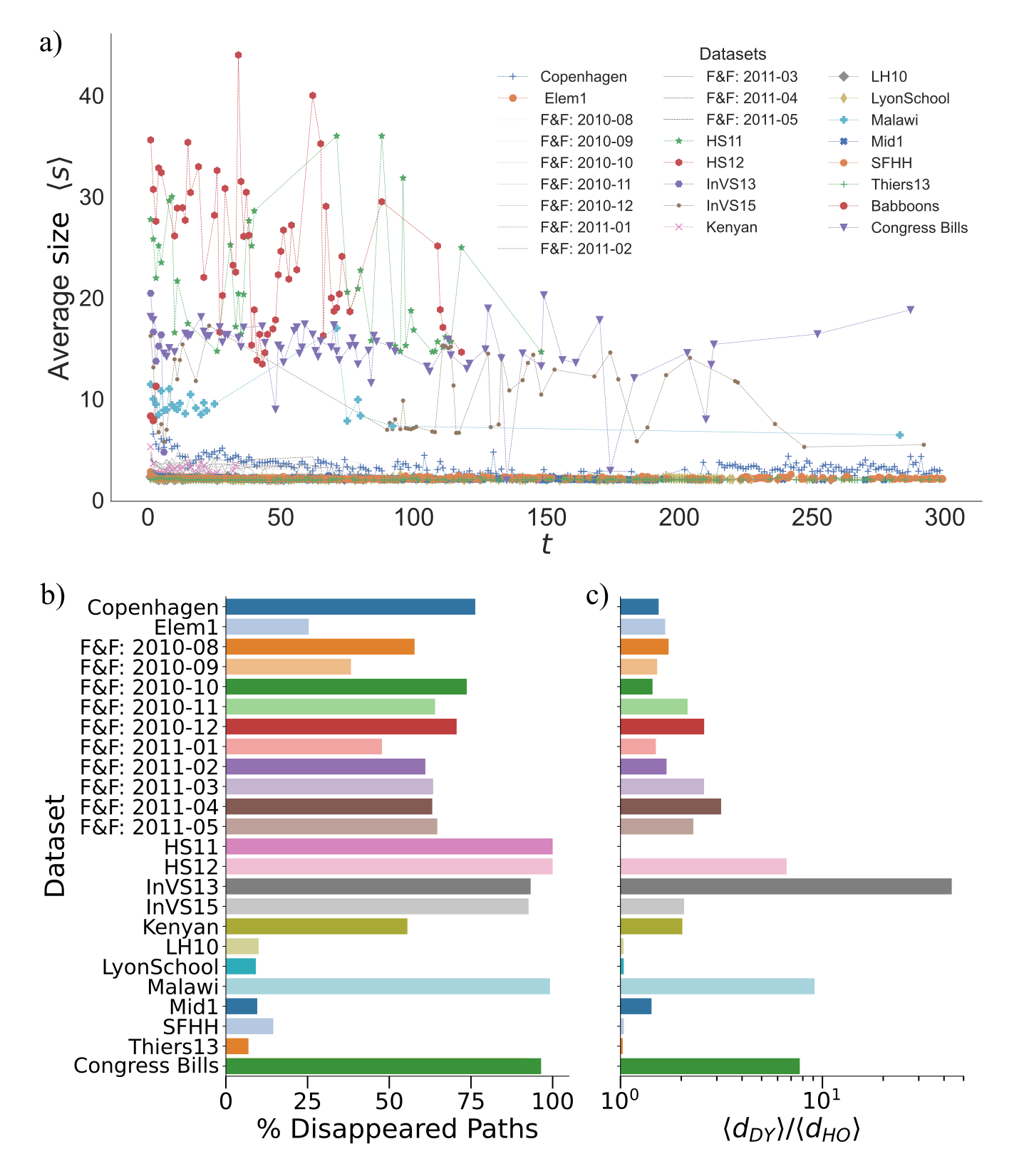}
  \end{subfigure}%
    \caption{ \justifying  
    Features of fastest paths in temporal hypergraphs. Average size $\langle{s}\rangle$ as a function of the path length across all datasets (a). 
    For each dataset, we also show the percentage of fastest paths that only exist as \ho paths (b), and how much longer in duration the dyadic paths are than the original \ho paths on average (c). 
    }
    \label{fig:6}
\end{figure}

\section*{CONCLUSION}

In this work we have unveiled the \ho organisation of shortest paths in systems with non-dyadic interactions \textcolor{black}{ and have shown the importance of non-dyadic ties for shortest paths.}
By investigating average path size in many real-world hypergraphs, \textcolor{black}{ we have seen that} \ho interactions are crucial to introduce shortcuts into the system, enabling efficient communications beyond system units that would otherwise be disconnected if only pairwise segments were considered. 
We extended our analysis to the case of temporal hypergraphs, investigating both temporally fastest and topologically shortest paths showing how real-world connectivity cannot be reproduced by simple randomised null models. 

\textcolor{black}{In the future it would be interesting to explore how the location and properties of hyperedges impact their contribution to shortest paths, which could be done using the notion of hypercoreness~\cite{mancastroppa2024hypercores}  or using more sophisticated reference models such as those that are known to preserve static shortest path lengths across randomisation~\cite{najajima2022randomising}.
Another idea to investigate whether
truly higher-order links are positioned better than pairwise links could involve counting segment duplication, extending the analysis in S4 of our Supplementary.}

\textcolor{black}{Higher-order connectivity can be defined more generally using for example hyperedge overlap~\cite{aksoy2020hypernetwork} or by extending notions of flow to the \ho case~\cite{estrada2008communicability}. A future avenue of exploration would be the analysis of our path size metric for networks where connectivity is defined more generally and an investigation of the changes that such definitions would induce~\cite{van2008observable}.}

Analysing large \ho temporal networks by calculating all shortest/fastest paths has a very high computational cost, 
as it requires global information about the system.
This makes the analysis particularly hard for very dense networks, such as the Copenhagen Network Study, or very large ones. In order to analyse larger systems, it would be important to develop more efficient methods, for instance based on the development of heuristics for \ho networks that would allow the calculation of shortest path lengths in an approximate but more efficient way, similar to what was recently done for the problem of \ho motifs~\cite{lotito2024exact}.  
{This would permit a clearer comparison also of the differences between shortest and fastest paths in temporal networks, an analysis that is currently limited by the computational infeasibility of calculating all temporal hyperpaths between 2 nodes. Future work could explore this avenue, considering for example the comparison of subsets of paths between nodes and future versions of themselves instead of between all nodes or other sampling schemes.}

Another possible extension of the work would be to investigate the effect of the size of the time window considered in our calculations on the emerging connectivity, as previously done for the case of burstiness~\cite{karsai2011small, cencetti2021temporal}. Finally, a study of \textcolor{black}{the contribution of hyperlinks of different orders to the shortest paths } might be used to develop alternative attack strategies to dismantle the system by affecting shortest paths.

Our work investigates the connectivity properties of static
and time-varying hypergraphs, contributing to a better understanding of the structural organisation of systems with \ho interactions.

\section*{METHODS}

\subsubsection*{Datasets}
This section provides more detailed descriptions of the \numdatasets datasets used in the above analyses. 
    
Thirteen of the datasets~\cite{sapiezynski2019copenhagen, aharony2011friendsnfamily, kiti2016quantifying, ozella2021malawi}
(\textit{Copenhagen, 
Friends \& Family (monthly between August 2010 - May 2011), 
Kenyan,  
Malawi}) 
describe generic face-to-face interactions between individuals. 
The Friends \& Family dataset is particularly rich as it spans an extended time period while simultaneously storing information at a very high time resolution. We therefore opt to split it into 10 smaller datasets, each tracking the temporal evolution over one calendar month. Additionally, following the procedure for intermediate aggregation as described below, we group time frames into blocks of 8 hours and create a static hypergraph for each interval. 

Six of the datasets~\cite{genois2018sociopatterns, fournet2014highschool, toth2015role} 
(\textit{LyonSchool and Thiers13, 
HS11 and HS12, 
Elem1 and Mid1}) 
describe interactions between pupils at school. 
Five datasets~\cite{genois2018sociopatterns,benson2018simplicial,fowler2006congressbills} describe interaction patterns within diverse work environments, namely a hospital 
(\textit{LH10, DAWN}), 
a workspace 
(\textit{InVS13 and InVS15}) and 
a conference (\textit{SFHH}).
For all of these datasets, nodes correspond to individuals and hyperedges correspond to group interactions.


The final dataset (\textit{Congress-bills}) describes political interactions between the US House of Representatives and Senate~\cite{fowler2006congressbills}. Nodes here represent congresspersons and each hyperedge contains all sponsors and co-sponsors of a specific legislative bill. 

 {In Section~5 of the Supplementary Materials, we additionally perform the same analyses on data describing coauthorship within geological journals across time, which is not inherently social but still a classic example of a setting in which \ho are natural.}

\subsubsection*{Higher-order reconstruction}
\label{sec:meth-horecon-aggr}

We use \numdatasets freely available datasets for our analyses describing group interactions. Even though interactions involve groups, the majority of the datasets store interactions between individuals as (time, node, node)-tuples as a computational simplification even when interactions involve groups. We are here interested in investigating the different contributions of \ho and dyadic edges to the \ho organisation of networks. To this end, we assume that if $k$ individuals are all pairwise connected at time $t$ (i.e. they form a clique), then together they form a group of size $k$ and are promoted to a hyperedge of size $k$ in the temporal hypergraph. 

 {Throughout our analyses, we distinguish between `pure' dyadic interactions and those which are an artefact of the data structures chosen to store the data digitally.} We consider an edge  {to be a `pure pairwise` interaction when} it does not form part of any larger clique, i.e. when it cannot be promoted to a group.  

We create the static hypergraph by omitting temporal information and generating a hypergraph with all nodes and interactions reported.  In effect, this collates all snapshots into a single network. This allows for potentially nested edges to exist in cases where a group of individuals at one time lose/gain members at an earlier/later time, even though in general we assume nested edges do not exist due to the way in which we create hyperedges.

\subsubsection*{ {Temporal aggregation in higher-order networks}}
To investigate the temporal evolution of social systems, we construct a temporal network consisting of a sequence of static hypergraphs, each associated with a specific timestamp. In some datasets we have very fine-grained temporal information with data logged at intervals of between 20s-5mins  {and which implies that many of the static hypergraphs are often extremely sparse. As a result, the temporal network will also be much more disconnected as nodes are active much less frequently. 
Following a procedure standard in the literature~\cite{masuda2016guide, holme2012temporal}, we address this by performing a preprocessing step of coarse-graining where we aggregate multiple smaller snapshots into a single bigger window (e.g. grouping together snapshots of resolution 30s into groups of 5 minutes).} The choice of aggregation window is determined in a data-driven manner by plotting the size of the largest connected component as a function of the size of the aggregation window and taking note of when its size saturates. We select the smallest window for which the largest connected component no longer grows.  

The duration of a temporal path from node $i$ at some time $t_1$ to node $j$ is defined as the time elapsed between time $t_1$ and the first time $t_n$ that node $j$ appears in a snapshot (i.e. becomes active) after $t_1$. 

In the case that multiple shortest/fastest paths exist between a given node-pair, we select a single path uniformly and at random. 
 {Whenever multiple hyperedges connect a single segment of a path, we select the hyperedge of minimum size and assign to the segment its size  {motivated by the intuition that smaller groups will more accurately transmit information as there are fewer places for information to flow towards~\cite{estrada2008communicability}. We more fully study the impact of choosing the minimum instead of alternatives in Section~3 of the Supplementary Materials.}

\subsubsection*{Calculating shortest paths in static hypergraphs}
Shortest paths in hypergraphs may be calculated by related them to the graph shortest path problem via a clique expansion of the hypergraph. 

Specifically, a shortest path between any node pair in a hypergraph is obtained by first transforming a static hypergraph to an undirected graph by placing a pairwise edge between 2 nodes whenever they share the same hyperedge (in effect, the clique expansion of the hypergraph). We then apply Dijkstra's shortest path algorithm~\cite{cormen2001section} to find the shortest paths in the pairwise graph. Finally, we assign to each pairwise link $(a,b)$ in the shortest path the associated hyperedge that containing that pair.

\subsubsection*{Calculating minimum-length paths}

Calculating fastest paths in temporal networks is a non-trivial task. To enable us to calculate time-respecting paths, we first map a temporal \ho network to a static digraph representation whose nodes are now tuples of (time, node) pairs from the original temporal hypergraph~\cite{valdano2015analytical, sato2019dyane}. We may then easily extract temporal paths from this new digraph. To calculate minimum-length topological paths, we assign a weight of 1 to all edges and employ Dijkstra's shortest path algorithm~\cite{cormen2001section}. To calculate minimum-duration temporal paths, we weight edges by the time interval traversed and again use Dijkstra's algorithm.  As before, in the presence of multiple paths we resolve ambiguities by picking one uniformly and at random. If a node `stores' a message across $t$ timesteps of length $dt$, this node is considered to contribute $t\cdot{dt}$ units to the temporal path length. 

Moreover, calculating shortest paths for large networks is computationally expensive, so we opt to pick a starting time $t_0$ and window length $w$ and label $V_{t_0}$ as the starting node-set. In our case, all datasets have a window size of 300 timesteps, except for the Friends \& Family dataset, which spans the entire calendar month. We then determine the existence and length of fastest and shortest paths from this starting set $V_{t_0}$ to all other nodes in the set $V$ within a timeframe of $w$ consecutive time snapshots. As before, non-existent paths are set to have a path length of $\infty$.


\subsubsection*{Randomisations}

We consider randomisations of the data in both the static and temporal settings to compare our results.
For both cases we generate 100 realisations of the particular null model and and plot the resulting averages and 97.5\% confidence intervals in the figures. 
For the static hypergraph, we use the \ho configuration model, an extension of the well-known configuration model that keeps the degree-distribution fixed across all orders and generates a new edge-set. 
For temporal hypernetworks, we use a randomisation that shuffles the timestamps of instantaneous events inside individual timelines  to create a rearranged version while still preserving both the underlying static hypergraph and the total number of events, following the methodology of~\cite{gauvin2022randomized}.

\section*{CODE AVAILABILITY}
The code to perform the analyses described will soon be publicly available as implemented functions within HypergraphX~\cite{lotito2023hypergraphx}, an open-source Python library for \ho network analysis.

\section*{DATA AVAILABILITY}
The datasets are public and their associated preprocessing will be made publicly available online at~\url{https://github.com/joanne-b-nortier/higher-order-shortest-paths}.

\section*{ACKNOWLEDGEMENTS}
F.B.~acknowledges support from the Austrian Science Fund~(FWF) through the project~10.55776/PAT1052824. 

\bibliographystyle{unsrt}
\bibliography{biblio}

\end{document}